\documentclass[preprint2]{aa}
\usepackage{lscape}
\usepackage{graphicx}
\usepackage{float}

\usepackage{titlesec}

\begin{document}


\title {
Seeds of Life in Space (SOLIS)\\
VIII. SiO isotopic fractionation and a new insight in the shocks of L1157-B1 \thanks{Based on observations carried out with the IRAM NOEMA interferometer. IRAM is supported by INSU/CNRS (France), MPG (Germany) and IGN (Spain)}
 \thanks{The cleaned cubes and the spectra are available in electronic form at the CDS via anonymous ftp to cdsarc.u-strasbg.fr (130.79.128.5) or via http://cdsweb.u-strasbg.fr/cgi-bin/qcat?J/A+A/}
}


\author{S. Spezzano\inst{1}  \and C. Codella\inst{2, 3}  \and L. Podio\inst{2}  \and C.Ceccarelli\inst{3}  \and P. Caselli\inst{1}  \and R. Neri\inst{4}  \and A. L\'opez-Sepulcre\inst{3,4}  }

\institute{Max Planck Institute for Extraterrestrial Physics, Giessenbachstrasse 1, 85748 Garching, Germany \and INAF, Osservatorio Astrofisico di Arcetri, Largo E. Fermi 5, 50125 Firenze, Italy \and Univ. Grenoble Alpes, CNRS, Institut de Plan\'etologie et d'Astrophysique de Grenoble (IPAG), 38000 Grenoble, France \and Institut de Radioastronomie Millim\'etrique, 300 rue de la Piscine, Domaine Universitaire de Grenoble, 38406, Saint-Martin d'H\`eres,
France}

\abstract {Contrary to what is expected from models of Galactic chemical evolution (GCE), the isotopic fractionation of silicon (Si) in the Galaxy has been recently found to be constant. This finding calls for new observations, also at cores scales, to re-evaluate the fractionation of Si.} 
{L1157-B1 is one of the outflow shocked regions along the blue-shifted outflow driven by the Class 0 protostar L1157-mm, and is an ideal laboratory to study the material ejected from the grains in very short timescales, i.e. its chemical composition is representative of the composition of the grains.}
{We imaged $^{28}$SiO, $^{29}$SiO and $^{30}$SiO $J$ = 2-1 emission towards L1157-B1 and B0 with the NOrthern Extended Millimeter Array (NOEMA) interferometer as part of the Seeds of Life in Space (SOLIS) large project. We present here a study of the isotopic fractionation of SiO towards L1157-B1. Furthermore, we use the high spectral resolution observations on the main isotopologue, $^{28}$SiO, to study the jet impact on the dense gas. We present here also single-dish observations obtained with the IRAM 30m telescope and Herschel-HIFI. We carried out a non-LTE analysis using a Large Velocity Gradient (LVG) code to model the single-dish observations.} 
{From our observations we can show that (i) the (2-1) transition of the main isotopologue is optically thick in L1157-B1 even at high velocities, and (ii) the [$^{29}$SiO/$^{30}$SiO] ratio is constant across the source, and consistent with the solar value of 1.5.}
{We report the first isotopic fractionation maps of SiO in a shocked region and show the absence of a mass dependent fractionation in $^{29}$Si and $^{30}$Si across L1157-B1. A high-velocity bullet in $^{28}$SiO has been identified, showing the signature of a jet impacting on the dense gas. With the dataset presented in this paper, both interferometric and single-dish, we were able to study in great detail the gas shocked at the B1a position and its surrounding gas.}

\titlerunning{SOLIS SiO in L1157-B1}
\maketitle

\section{Introduction}
Studying the early stages in star and planetary systems formation is important for understanding the evolution of matter in the ISM, and finally our astrochemical origins. In particular, isotopic ratios are of pivotal importance to follow the chemical link in the evolution of the material in dense molecular clouds, nursery of new stars, to a planetary system like our own (e.g. Caselli \& Ceccarelli 2012, and references therein).

Jets are common in protostars, as they are an efficient way to release angular momentum (\citealt{frank14}, and references therein), and they have a deep effect on the chemical and physical evolution of the ISM. Indeed, shocks created by the fast (at least 100 km s$^{-1}$) jet propagating through the protostellar high-density cocoon are a unique place to study what is released by dust mantles and refractory cores as a consequence of processes such as sputtering and shattering \citep{bachiller01,fontani14,codella17, lefloch17,  cheng19, kong19,ospina-zamudio19, tychoniec19}.

\begin{figure*}
\centering
 \includegraphics [width=1.0\textwidth]{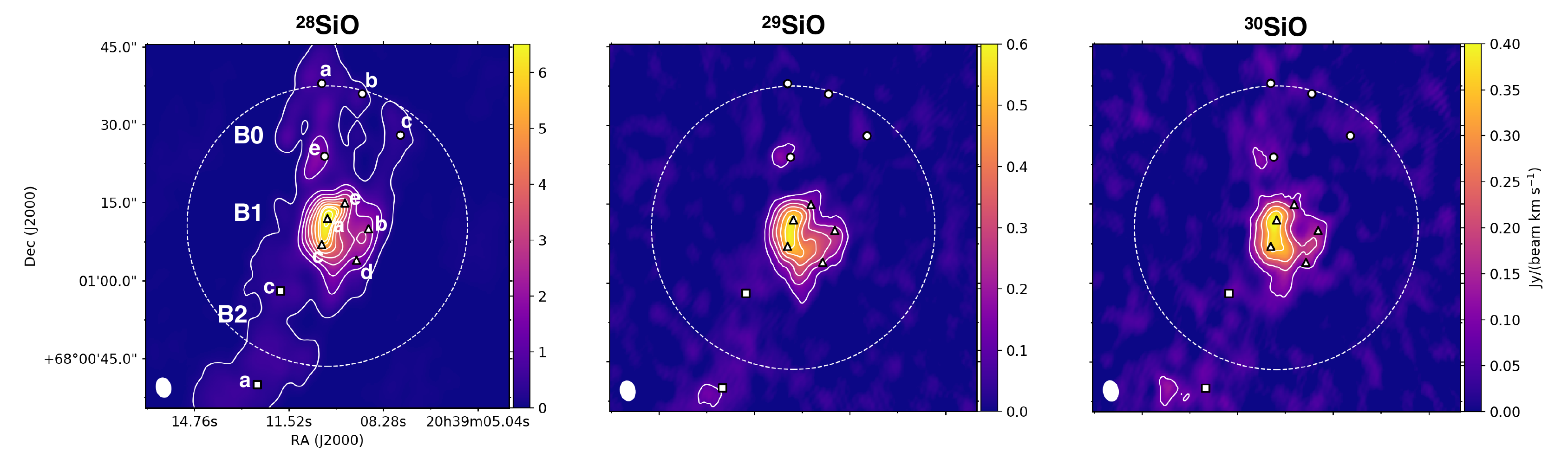}
 \caption{ $^{28}$SiO, $^{29}$SiO and $^{30}$SiO (2-1) integrated intensity maps towards L1157-B1, obtained with $\sim$7 km s$^{-1}$ velocity resolution. The emission has been integrated between -22.95 and 6.25 km s$^{-1}$. The clean beam of 3.8$\arcsec$ $\times$ 2.8$\arcsec$ is shown at the bottom left of the maps, and the primary beam of 56$\arcsec$ is reported as a dotted white line. The contours for $^{28}$SiO start at 5$\sigma$ (rms = 36 mJy/beam km s$^{-1}$), then 30$\sigma$, and then increase with steps of 20$\sigma$. The contours for $^{29}$SiO start at 5$\sigma$ (rms = 18 mJy/beam km s$^{-1}$), and increase with steps of 5$\sigma$. The contours for $^{30}$SiO start at 5$\sigma$ (rms = 17 mJy/beam km s$^{-1}$), and increase with steps of 5$\sigma$. B0, B1 and B2 are three shock regions along the outflow of L1157, and they are composed by several sub-clumps marked here with circles (B0), triangles (B1), and squares (B2) \citep{benedettini07}. The L1157-mm protostar is located towards the north-west at $\Delta\alpha$ = -25\arcsec and $\Delta\delta$ = +63.5\arcsec ($\alpha _{2000}$ = 20$^h$39$^m$06$^s$.0,  $\delta _{2000}$ = +68$^\circ$02$'$14$''$.0). The maps are not primary beam corrected.
}
  \label{fig:integrated-int}
\end{figure*}

One prototypical shock laboratory is the L1157 outflow.
The Class 0 low-mass protostar L1157-mm, located at 350 pc \citep{Zucker19}, is characterised by an episodic precessing jet \citep{gueth96,podio16} that produced several shocked regions. L1157-B1 is a shocked region along the outflow driven by L1157-mm. L1157-B1 is one of the best places to study the material released from dust mantles because it is kinematically very young ($\leqslant$2$\times$10$^3$ yr) \citep{codella17}. 
Previous observations of L1157-B1 have shown a very rich chemistry \citep{bachiller01,arce08,codella09, fontani14,codella15, codella17, lefloch17, codella20}. Interferometric observations show structure at small scale along the southern outflow lobe of L1157: the three large molecular clumps, called B0, B1 and B2, have been resolved into more subclumps \citep{benedettini07}. Some of the subclumps in B0, B1 and B2 are shown in Figure~\ref{fig:integrated-int}.
Based on the jet radial velocity observed in the inner knots and on the precession model presented in \cite{podio16}, the kinematical ages of the different shocked regions range from 1340 yr (B0), to 1550 yr (B1a) and 2530 yr (B2), at a distance of 350 pc.
In particular, L1157-B1 is a laboratory to study the molecules released in the gas-phase from the surface of dust grains, and the interplay of the chemistry in the gas-phase and on the dust.

SiO is an excellent tracer of shocked gas in outflows. Silicon is released in the gas phase from the grains because of sputtering and grain-grain collisions along the shock, and it is oxidised to SiO \citep{schilke97,caselli97}. Therefore, SiO can be studied efficiently in a shock given that its abundance is expected to increase by orders of magnitude (\citealt{gusdorf08a,gusdorf08b,guillet11,podio17}, and references therein).
The studies of SiO towards L1157-B1 include: the interferometric map of the 2-1 transition of the main isotopologue \citep{gueth98}, single-dish observations of multiple lines of the main isotopologue \citep{bachiller97, bachiller01, nisini07}, and single-dish observations of the rarer isotopologues $^{29}$SiO and $^{30}$SiO \citep{podio17}. 

Here we present the first maps of the silicon isotopologues of SiO towards L1157-B1 ($^{28}$SiO, $^{29}$SiO and $^{30}$SiO) observed in the framework of SOLIS (Seeds Of Life In Space; Ceccarelli et al. 2017), an IRAM NOrthern Extended Millimeter Array (NOEMA) large programme aimed at investigating the formation of complex molecules during the early stages of the star formation process.
With this dataset we can study the isotopic fractionation of Si around a Sun-like protostar, and better constrain the origin of the SiO high velocity component.
Furthermore, we complement the SOLIS data with previously obtained single-dish observations of SiO emission towards L1157-B1 from J=2-1 to 12-11.

The paper is structured as follows: the observations carried out with NOEMA, the IRAM 30m telescope and Herschel-HIFI are presented in Section 2, the results of the single-dish and interferometric observations are presented in Section 3, with a focus on the SiO isotopic fractionation in L1157-B1 and the kinematics. The non-LTE analysis of the IRAM 30m spectra are summarised in Section 4. In Section 5 we discuss the isotopic fractionation of SiO in the framework of the Galactic chemical evolution, as well as the signature of the jet coming from the protostar L1157-mm. Finally, our conclusions are summarised in Section 6.

\begin{figure}
\centering
 \includegraphics [width=0.4\textwidth]{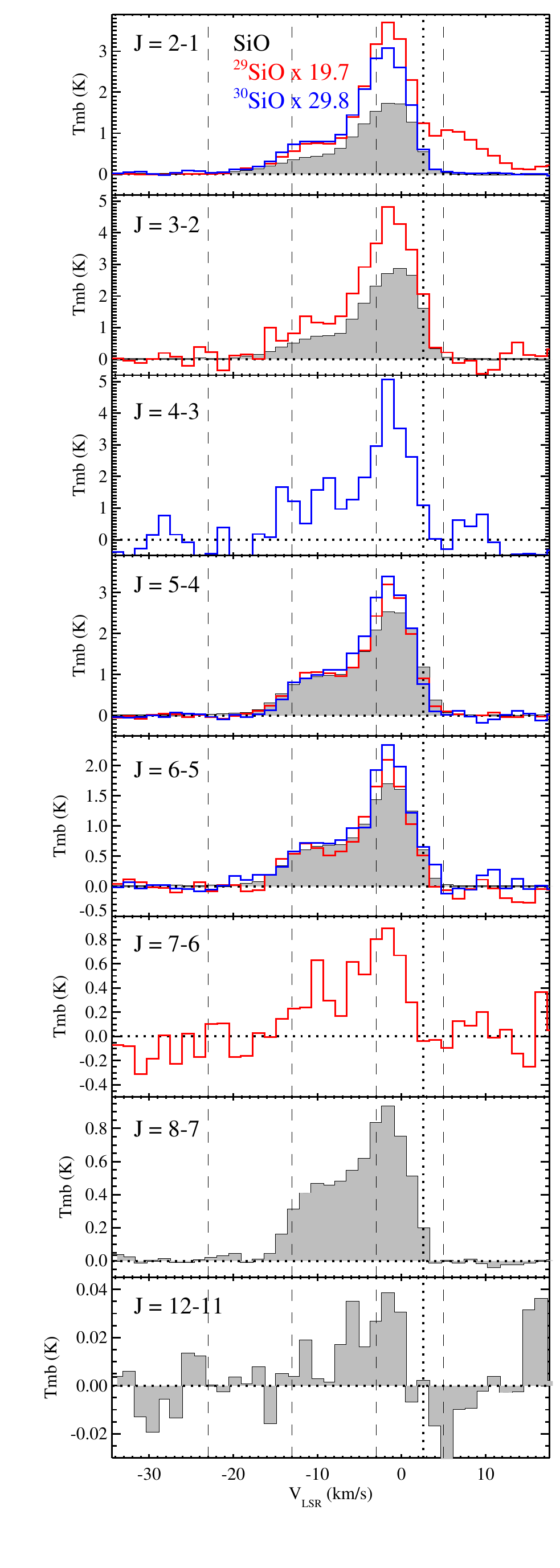}
\caption{Line profiles of SiO  in the L1157-B1 shock. The  line intensity in main beam temperature (T$_{\rm mb}$) from $^{29}$SiO (red), and $^{30}$SiO (blue) is over-plotted on the main isotopologue, $^{28}$SiO (grey), by multiplying for their Solar isotopic ratios ([$^{28}$Si/$^{29}$Si]$_{\odot}$ =19.7, [$^{28}$Si/$^{30}$Si]$_{\odot}$ = 29.8, \citealt{solar}). The transitions are labeled. The baseline and the systemic velocity (+ 2.6 km s$^{-1}$, \citealt{bachiller01}) are indicated by the horizontal and vertical dotted lines, respectively. The vertical dashed lines indicate the three velocity intervals where the integrated intensity has been calculated, namely [-23, -13],    [-13, -3] and  [-3, +5] km s$^{-1}$, see Table 1. The red-shifted component which seems to be associated to $^{29}$SiO $2-1$ is due to blending with the $^{13}$CH$_3$OH 4$_{1,3}$-3$_{-2,2}$ line.}
  \label{fig:spectra}
\end{figure}

\section{Observations}
\subsection{NOEMA}
L1157-B1 was observed for 6.7 h at 3 mm with the IRAM NOEMA array during 1 track on April 7th
2018 (9 antennas) using the C configuration. The shortest and longest projected baselines are 24 m and 644 m, respectively. The phase
centre is $\alpha _{2000}$ = 20$^h$39$^m$10$^s$.21,  $\delta _{2000}$ = +68$^\circ$01$'$10$''$.5.
The $^{28}$SiO, $^{29}$SiO and $^{30}$SiO $J$ = 2-1 lines listed
in Table 1 were observed using the Polyfix correlator, with a total
spectral band of  $\sim$8 GHz, and a spectral resolution of 2.0
MHz (7.3 km s$^{-1}$). The main isotopologue has been observed
also at high spectral resolution (62.5kHz, 0.22 km s$^{-1}$). Calibration was carried out following standard
procedures, using GILDAS-CLIC \footnotemark. The bandpass was calibrated on 3C84, while the absolute flux was fixed by observing LkH$\alpha$101, 2010+723, and 1928+738. The final uncertainty on the absolute flux scale is $\leqslant$10$\%$. The phase rms
was $\leqslant$60$^\circ$, the typical precipitable water vapour (pwv) was $\sim$ 6-10 mm, and the system temperature $\sim$ 80-120 K. Images were produced using natural weighting, and restored with a clean beam of 3.$''$8 $\times$ 2.$''$8 (PA= --167.8$^\circ$). The rms noise in the broadband cubes was 0.6-1.7 mJy beam$^{-1}$, depending on the frequency.
Our NOEMA observations recover 80$\%$ of the flux from the $^{28}$SiO (2-1) emission compared to the IRAM 30m observations, see Appendix B. The integrated
intensities of the rarer isotopologues in both the NOEMA and IRAM 30m spectra match within
20\%. Being 20\% consistent with the calibration error of the observations, it does not have an impact on the quantities that we derive in the isotopic ratio maps.

\subsection{IRAM 30-m}
The single-dish observations of the SiO, $^{29}$SiO, and $^{30}$SiO  lines from $J$ = 2-1 to $J$ = 8-7 (see Table 1 for the frequencies) have been obtained using the IRAM 30-m antenna in the context of the ASAI Large Program (Astrochemical Surveys At IRAM; Lefloch et al. 2018). The spectra are shown in Figure~\ref{fig:spectra}. Observations have been done in Wobbler Mode at the position of the B1 shock , namely at $\alpha_{\rm 2000}$ = 20$^h$ 39$^m$ 10$\farcs$2, $\delta_{\rm 2000}$ = +68$\degr$ 01$\arcmin$ 10$\farcs$5, i.e. at $\Delta$ = +25$\arcsec$ and $\Delta$ = -63$\farcs$5 from the L1157-mm protostar. The present observations have been performed during several runs in 2011 and 2012 using the EMIR receiver connected with the FTS spectrometers providing a spectral resolution of 200 kHz (i.e. 0.2 km s$^{-1}$ to 0.7 km s$^{-1}$, depending on the frequency). More details can be found in \cite{lefloch18}. The weaker spectra have been smoothed to 1.4 km s$^{-1}$ to increase the S/N ratio. The Half Power Beam Width (HPBW) spans from 7$\arcsec$ to 29$\arcsec$, depending on the frequencies. The forward (F$_{eff}$) and beam (B$_{eff}$) efficiencies are in the 0.80-0.95, and 0.34-0-81 ranges, respectively. The final rms, in main-beam temperature, goes from 1 mK (at 85 GHz) to 25 mK (342 GHz).

\subsection{Herschel-HIFI}
The SiO, $^{29}$SiO, and $^{30}$SiO $J$ = 12-11 transitions (Table 1 reports their frequencies) were observed with Herschel-HIFI on 2010 October 27 and 2009 August 1, during the unbiased spectral survey CHESS (e.g. Ceccarelli et al. 2010) with the HIFI bands 1a, at the same position used for the IRAM 30-m observations. The spectra are shown in Figure~\ref{fig:spectra}. A receiver in double side band mode was used, with a total integration time of 8912 s. The Wide Band Spectrometer was used, with a velocity resolution of about 0.15 km s$^{-1}$. The SiO(12-11) spectrum has been successively smoothed to 4 km s$^{-1}$ to increase the S/N ratio. The forward (F$_{eff}$) and beam (B$_{eff}$) efficiencies are 0.96 and 0.73, respectively. The Half Power Beam Width (HPBW) is about 42$\arcsec$ \citep{roelfsema12}. Data were processed with the ESA-supported package HIPE (Herschel Interactive Processing Environment) and analysed using the GILDAS-CLASS software \footnotemark[\value{footnote}]. The flux uncertainty is around 15\%. The rms, in main-beam temperature, is about 12-13 mK.

\footnotetext{http://www.iram.fr/IRAMFR/GILDAS/}

\begin{table*}
  \caption[]{\label{tab:lines} Properties of the detected transitions from SiO, $^{29}$SiO, and $^{30}$SiO in L1157-B1 (single-dish ASAI+CHESS spectra).}
    \scalebox{0.9}{
  \begin{tabular}[h]{cccccccccc}
    \hline
    \hline
line &  Frequency\tablefootmark{a} & E$_{\rm up}$ & HPBW & rms  &  V$_{\rm peak}$ &  T$_{\rm mb-peak}$ & $\int{T_{\rm mb} dV}$ [-23, -13] &   [-13, -3] &  [-3, +5]\\ 
       & MHz  & (K)             & (")       & (mK) &  (km/s)           &   (K)                      & (K km/s)                                    & (K km/s)   & (K km/s) \\
\hline 
\multicolumn{10}{c}{SiO}\\
 2-1 &  86846.96 &    6 &   28 &    2 &  -1.6 & 1.72 &  1.51 $\pm$  0.01 &  8.22 $\pm$  0.01 &  9.60 $\pm$  0.01 \\ 
 3-2 & 130268.61 &   13 &   19 &    9 &  -0.2\tablefootmark{b} & 2.86 &  1.93 $\pm$  0.03 & 12.13 $\pm$  0.03 & 17.41 $\pm$  0.03 \\ 
 5-4 & 217104.98 &   31 &   11 &    3 &  -1.6 & 2.51 &  2.58 $\pm$  0.01 & 13.03 $\pm$  0.01 & 15.11 $\pm$  0.01 \\ 
 6-5 & 260518.02 &   44 &    9 &    5 &  -1.6 & 1.69 &  1.66 $\pm$  0.02 &  8.93 $\pm$  0.02 &  9.37 $\pm$  0.02 \\ 
 8-7 & 347330.59 &   75 &    7 &   17 &  -1.6 & 0.93 &  0.83 $\pm$  0.06 &  5.75 $\pm$  0.06 &  4.49 $\pm$  0.06 \\ 
 12-11$^{c}$ & 520881.09 &163 &   41 &   13 &  -1.6 & 0.04 &  $<0.05$          &  0.17 $\pm$  0.05 &  0.06 $\pm$  0.04 \\ 
\hline 
\multicolumn{10}{c}{$^{29}$SiO}\\
 2-1 &  85759.20 &    6 &   29 &    1 &  -1.6 & 0.19 & 0.12 $\pm$ 0.01 & 0.74 $\pm$ 0.01 & -\tablefootmark{c} \\ 
 3-2 & 128637.05 &   12 &   19 &   10 &  -1.6 & 0.24 & 0.18 $\pm$ 0.04 & 1.03 $\pm$ 0.04 & 1.34 $\pm$ 0.03 \\ 
 5-4 & 214385.75 &   31 &   11 &    3 &  -1.6 & 0.16 & 0.11 $\pm$ 0.01 & 0.72 $\pm$ 0.01 & 0.83 $\pm$ 0.01 \\ 
 6-5 & 257255.22 &   43 &   10 &    4 &  -1.6 & 0.11 & 0.07 $\pm$ 0.02 & 0.46 $\pm$ 0.02 & 0.49 $\pm$ 0.01 \\ 
 7-6 & 300120.47 &   58 &    8 &   10 &  -1.6 & 0.05 & $<0.04$         & 0.25 $\pm$ 0.04 & 0.18 $\pm$ 0.03 \\ 
 8-7 & 342980.84 &   74 &    7 &   25 &    -  &  -   & $<0.1$          & $<0.1$          & $<0.1$                \\
 12-11$^{c}$ & 514359.34 &161 &   41 &   13 &    -  &  -   & $<0.05$         & $<0.05$         & $<0.05$               \\  
\hline 
\multicolumn{10}{c}{$^{30}$SiO}\\
 2-1 &  84746.17 &    6 &   29 &    1 &  -1.6 & 0.10 &  0.10 $\pm$  0.01 &  0.49 $\pm$  0.01 &  0.52 $\pm$  0.01 \\ 
 4-3 & 169486.88 &   20 &   15 &   28 &  -1.6 & 0.17 &  $<0.1$           &  0.6  $\pm$  0.1  &  0.7  $\pm$  0.1 \\ 
 5-4 & 211853.47 &   31 &   12 &    3 &  -1.6 & 0.11 &  0.06 $\pm$  0.01 &  0.53 $\pm$  0.01 &  0.57 $\pm$  0.01 \\ 
 6-5 & 254216.66 &   43 &   10 &    3 &  -1.6 & 0.08 &  0.07 $\pm$  0.01 &  0.35 $\pm$  0.01 &  0.39 $\pm$  0.01 \\ 
 12-11\tablefootmark{d} & 508285.78&  159 &   42 &   12 &  -    &  -   &  $<0.05$          &  $<0.05$          & $<0.05$ \\ 
    \hline     
  \end{tabular}
  }
 \tablefoot{
\tablefoottext{a}{\cite{muller13}}
\tablefoottext{b}{The velocity difference between the 3-2 $^{28}$SiO transition and all other transitions reported here is only one velocity channel, hence we consider its velocity in accordance with the others.}
\tablefoottext{c}{ The integrated intensity at low velocities of the $2-1$ transition of $^{29}$SiO is not reported here because the line is blended with the 4$_{1,3}$-3$_{-2,2}$ transition of $^{13}$CH$_3$OH.}
\tablefoottext{d}{ The $12-11$ transitions are taken with {\it Herschel}/HIFI (LP CHESS). All the other spectra are taken with IRAM-30m (LP ASAI).}}
\end{table*}

\section{Results}
\subsection{Single-dish spectra}
We detected 6 lines of $^{28}$SiO (up to $J$ = 12-11), 5 lines of $^{29}$SiO (up to $J$ = 7-6),
and 4 lines of $^{30}$SiO (up to $J$ = 6-5), thus collecting a large SiO dataset (spanning from 86 GHz to 521 GHz). 
The spectra are shown in Figure~\ref{fig:spectra} and the line properties are summarised in Table 1.
The SiO lines peak at blue-shifted velocity, $\sim$ --4 km s$^{-1}$ with respect to systemic velocity (v$_{sys}$ = +2.6 km s$^{-1}$). Moreover, the SiO lines show a high-velocity wing extending up to -20 km s$^{-1}$ likely originating in the strong shock occurring where the jet impacts the cavity walls (see Section 3.4). The single-dish spectra are used to characterise the physical properties of the emitting gas via the LVG analysis described in Section~\ref{sec:lvg}, and therefore the integrated intensities reported in Table 1 are divided into three velocity intervals: low velocity (LV) from -3 to +5 km s$^{-1}$, high velocity (HV) from -13 to -3 km s$^{-1}$, and extreme high velocity (EHV) from -23 to -13 km s$^{-1}$.

\subsection{Integrated intensity maps}
Figure~\ref{fig:integrated-int} shows the image of the emission of the $J$=2-1 transitions of $^{28}$SiO, $^{29}$SiO and $^{30}$SiO obtained with $\sim$7 km s$^{-1}$ spectral resolution. 
The intensity has been integrated between -22.95 and 6.25 km s$^{-1}$, i.e. in 4 channels centred at 2.6, -4.7, -12, and -19.3 km s$^{-1}$ respectively, where emission from the rarer SiO isotopologues is detected (see Section \ref{sec-fractionation}). The channel maps for all three isotopologues are presented in the Appendix A (Figures A.1-A.3).
The map of $^{28}$SiO (2-1) had already been reported in \cite{gueth98} using
IRAM-PdBI with 4 antennas. The present NOEMA dataset has an improvement in the noise level of a factor of 20 (rms = 18.3 mJy/beam vs. 0.9 mJy/beam) and allows us to image higher (blue) velocities, 
up to --20 km s$^{-1}$, as described in more detail  in Section \ref{jet}.
The integrated intensity maps in Figure~\ref{fig:integrated-int} show the structure of the bow shock B1, the B0 wall, and also some structure towards the shocked region B2. The rarer isotopologues show
the same emission morphology of the main isotopologue, $^{28}$SiO, with the maximum located at $\alpha _{2000}$ = 20$^h$39$^m$10$^s$.26,  $\delta _{2000}$ = +68$^\circ$01$'$10$''$.61. This position is consistent, within the beam size, with the B1a clump shown in \cite{benedettini07} and thought to be the position where the precessing jet driven by L1157-mm impacts the cavity wall \citep{gueth98}.
The rarer isotopologues also trace the so-called "finger" in the south part of B1, identified at low velocities in SiO \citep{gueth98} and in CS \citep{benedettini07}, which lies along the jet direction and is thought to be associated with the magnetic precursor of the shock \citep{bachiller01}.
Given that B2 is outside the primary beam, we will be cautious on quantitative measurements towards that position.

\subsection{SiO isotopic fractionation in L1157-B1}\label{sec-fractionation}
\begin{figure}
\centering
 \includegraphics [width=0.5\textwidth]{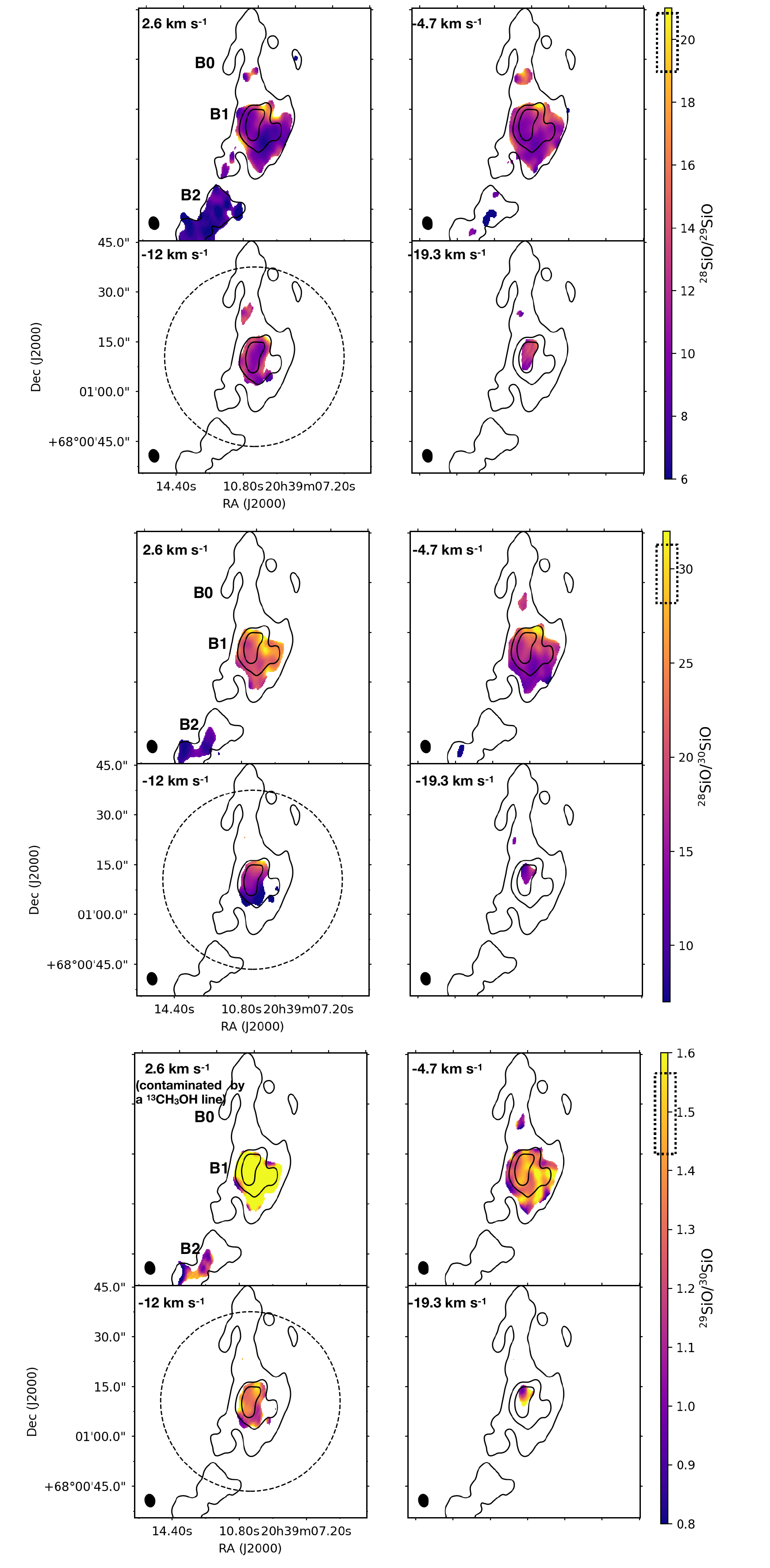}
 \caption{(top) $^{28}$SiO/$^{29}$SiO, (middle) $^{28}$SiO/$^{30}$SiO, and (bottom) $^{29}$SiO/$^{30}$SiO isotopic ratio channel maps. The clean beam of 3.8$\arcsec$ $\times$ 2.8$\arcsec$ is shown at the bottom left of the maps, and the primary beam of 56$\arcsec$ is reported as a dotted black circle. The contours show the 10, 60 and 120 $\sigma$ levels of the integrated intensity map of the $^{28}$SiO, showed in Figure 1. The Solar isotopic ratio of [$^{28}$Si/$^{29}$Si] = 19.7, [$^{28}$Si/$^{30}$Si]= 29.8, and [$^{29}$Si/$^{30}$Si]= 1.5, including a 5$\%$ error, is reported as a dotted box in the colour bar \citep{solar}. 
}
  \label{fig:isotopic_ratio}
\end{figure}

While the emission of the main isotopologue, $^{28}$SiO, is detected with more than 5$\sigma$ in all six channels shown in Figure A.1, the rarer isotopologues are detected only in the central 4 channels, from +2.6 to --19.3 km s$^{-1}$, shown in Figures A.2 and A.3. For this reason, we only use these channels to compute the integrated intensity maps (Figure~\ref{fig:integrated-int}) and the isotopic ratio maps (Figure~\ref{fig:isotopic_ratio}). 
A threshold of 5$\sigma$ has been applied to the maps of the different isotopologues before computing the isotopic ratio maps.
The Solar isotopic ratios are: [$^{28}$Si/$^{29}$Si] = 19.7,  [$^{28}$Si/$^{30}$Si]= 29.8, and [$^{29}$Si/$^{30}$Si]= 1.5 \citep{solar}. 

The [$^{28}$SiO/$^{29}$SiO] and [$^{28}$SiO/$^{30}$SiO] ratios in the maps shown in Figure~\ref{fig:isotopic_ratio} (upper and middle panel) are lower than the Solar ratios (shown as a dotted box in the colour bar), which can be interpreted as an indirect proof that the $^{28}$SiO (2-1) line is optically thick in the whole region and at all velocities that we imaged. 
\cite{podio17} measured the [$^{28}$SiO/$^{29}$SiO] and [$^{28}$SiO/$^{30}$SiO] ratios from single-dish observations of the (5-4) and (6-5) transitions, and showed that while the high-velocity wings are in agreement with the solar isotopic ratios, in the low velocities the ratios are 10$\%$--15$\%$ lower, suggesting a large optical depth of the main isotopologue.

Contrary to what happens in the isotopic ratio maps involving the main isotopologue, the [$^{29}$SiO/$^{30}$SiO] ratio shown in the lower panel of Figure~\ref{fig:isotopic_ratio} is quite uniform around the Solar ratio of 1.5 in all channels except the one at +2.6 km s$^{-1}$. This deviation can be explained with a contamination by the 4$_{1,3}$-3$_{-2,2}$ transition of $^{13}$CH$_3$OH, as shown in Figure 2 of \cite{podio17}. 
The value of the [$^{29}$SiO/$^{30}$SiO] ratio averaged within a beam in the channel at --19.3 km s$^{-1}$ is 1.14, while in the channels at --12.0 km s$^{-1}$ and --4.7 km s$^{-1}$ ranges from 1.1 to 1.5. Considering an uncertainty of $\sim$20$\%$ on our maps and of $\sim$5$\%$ on the Solar isotopic ratio \citep{solar}, these values are consistent with the Solar value of 1.5, and do not show any variation across the source.
Studying the isotopic fractionation in galactic sources (as L1157-B1), and
comparing it with the Solar value (defined 4.6 Gyr ago), gives information on the environment where the Sun and Solar system were formed, as well as the processes that formed them. Our results confirm what was reported in \cite{podio17} based on single-dish observations: there has not been a significant change in the Si isotopic ratios in the past 4.6 billion years. Furthermore, with the present dataset, we can now exclude local changes in the isotopic ratios within the shocked region in L1157-B1.

\begin{figure}
\centering
 \includegraphics [width=0.5\textwidth]{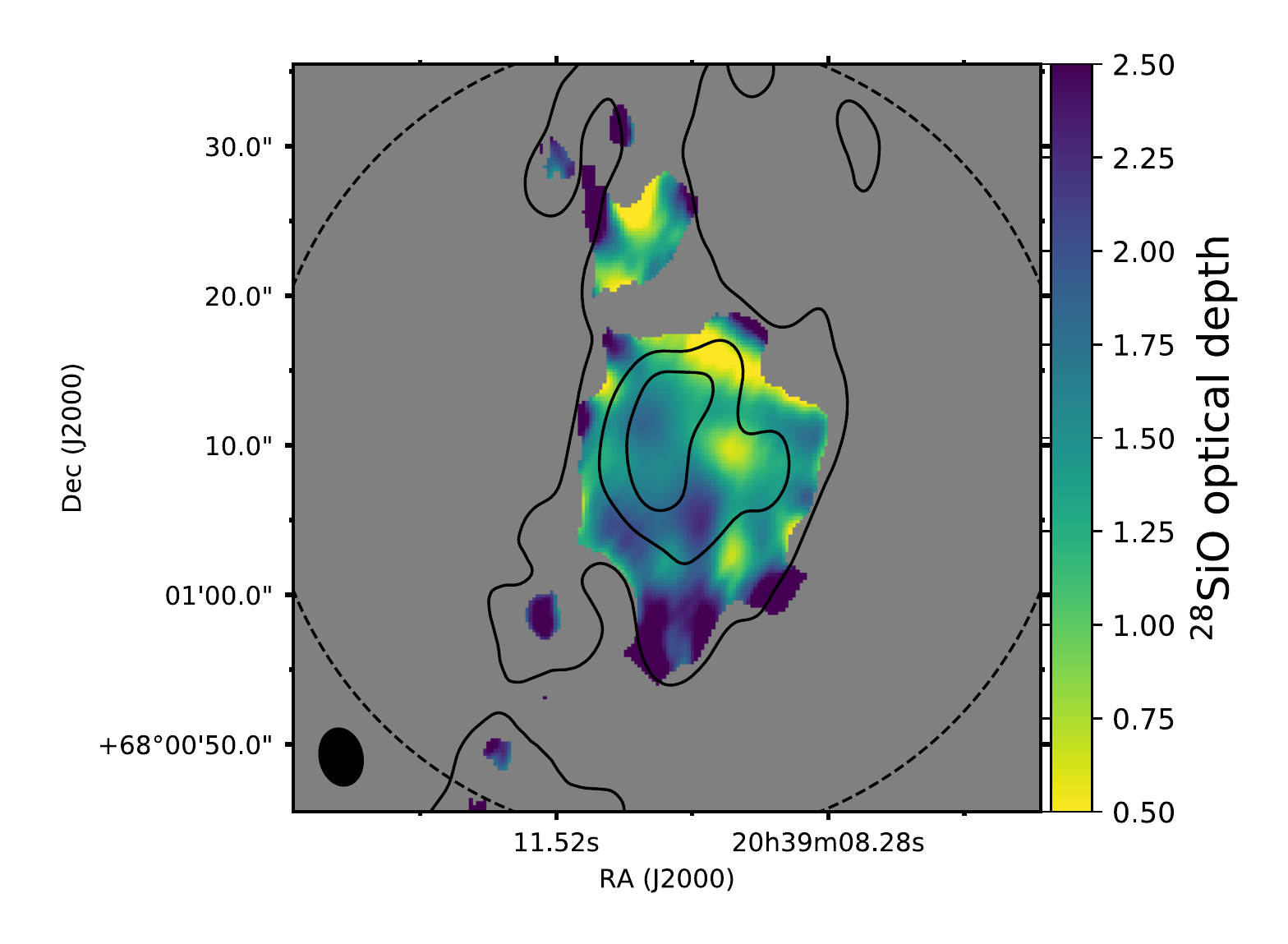}
 \caption{$^{28}$SiO (2-1) optical depth map computed from the $^{28}$SiO/$^{30}$SiO ratio assuming the solar isotopic ratio [$^{28}$Si/$^{30}$Si]= 29.8. The contours show the 10, 60 and 120 $\sigma$ levels of the integrated intensity map of the $^{28}$SiO, showed in Figure~\ref{fig:integrated-int}. The primary beam of 56$\arcsec$ is reported as a dotted black circle. 
}
  \label{fig:tau}
\end{figure}

We compute the optical depth ($\tau$) map for the $^{28}$SiO (2-1) transition across the core, see Figure~\ref{fig:tau}. Following what is described in \cite{mangum15}, we have computed the optical depth of the main isotopologue assuming that the emission of both $^{28}$SiO and $^{30}$SiO is co-spatial, that the atomic Solar ratio of [$^{28}$Si/$^{30}$Si]= 29.8 is the same as the molecular ratio, and that the two isotopologues have similar line profiles:

\begin{eqnarray}
\frac{\int T_R(^{28}SiO) dV}{\int T_R(^{30}SiO) dV} &=& \frac{ 1- exp^{-\tau({^{28}SiO)}}}{1- exp^{-29.8\tau({^{28}SiO)}}}.
\end{eqnarray}

Towards the B0 and B1 clumps, the values of $\tau$ averaged within a beam range from 1.0 to 2.2, comparable with the values reported in \cite{podio17} derived from single-dish observations ($\tau$ = 0.7-1.9). The southern clump B2 is outside the primary beam and hence not considered in this discussion.

\subsection{Kinematics: the jet signature}\label{jet}
\begin{figure*}
\centering
 \includegraphics [width=0.8\textwidth]{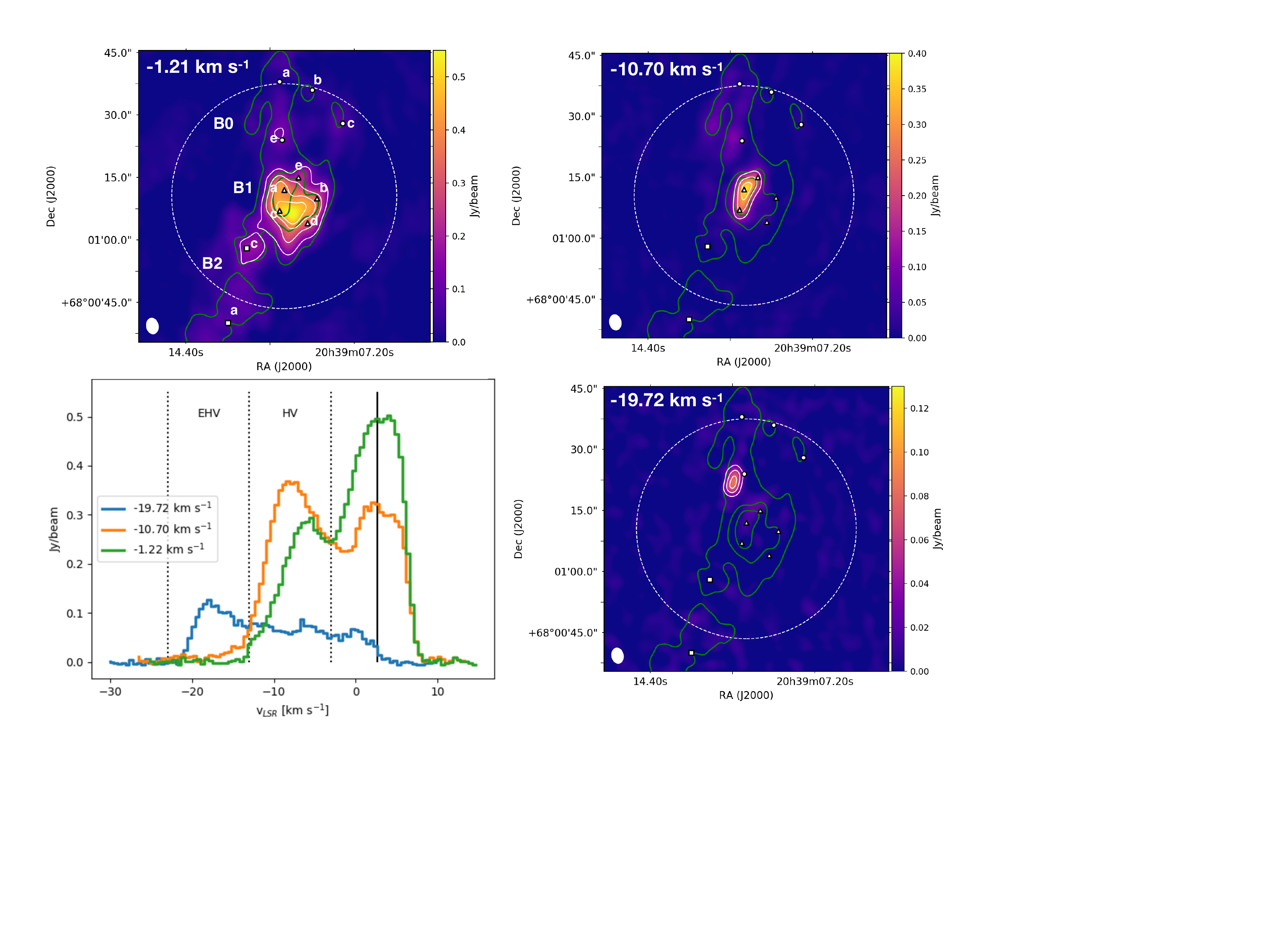}
 \caption{Velocity channel maps of $^{28}$SiO observed with high spectral resolution (0.5 km s$^{-1}$) at very high (-19.72 km s$^{-1}$), high (-10.70 km s$^{-1}$), and low velocity (-1.22 km s$^{-1}$), and the spectra (bottom left) extracted in the position of intensity peak of each channel. 
The white contours in the -19.72 km s$^{-1}$ channel map start at 5$\sigma$, with steps of 5$\sigma$ (rms = 5 mJy beam$^{-1}$).
The white contours in the -10.70 km s$^{-1}$ channel map start at 25$\sigma$, with steps of 20$\sigma$.
The white contours in the -1.22 km s$^{-1}$ channel map start at 25$\sigma$, with steps of 25$\sigma$.
The clean beam of 3.8$\arcsec$ $\times$ 2.8$\arcsec$ is shown at the bottom left of the maps, and the primary beam of 56$\arcsec$ is reported as a dashed white circle. The green contours in the channel maps show the 10, 60 and 120 $\sigma$ levels of the integrated intensity map of the $^{28}$SiO, showed in Figure~\ref{fig:integrated-int}.  B0, B1 and B2 are three shock regions along the outflow of L1157, and they are composed by several sub-clumps marked here with circles (B0), triangles (B1), and squares (B2) \citep{benedettini07}. The vertical solid line in the spectra shows the systemic velocity of L1157-B1, +2.6 km s$^{-1}$. The vertical dotted lines show the intervals of velocity used to run the LVG, (-23, -13) km/s for the extremely high velocity component (EHV) and (-13, -3) km/s for the high velocity component (HV).
}
  \label{fig:jet_signature}
\end{figure*}

As reported in Section 2, we have imaged the main isotopologue, $^{28}$SiO, also with high-velocity resolution (0.22 km s$^{-1}$). We successively smoothed the spectra to 0.5 km s$^{-1}$ in order to further increase the S/N ratio (see Figure 5 for a sample of spectra), and obtained an unprecedented view of the high velocity component of the (2-1) transition. This allows us to look for signatures of the expected impacts of the jet against the cavity walls.
In Figure~\ref{fig:jet_signature}, we show three representative velocity channel maps, at very high (--19.72 km s$^{-1}$), high (--10.70 km s$^{-1}$), and low velocity (--1.22 km s$^{-1}$): using such images we selected the intensity peaks and extracted the corresponding spectra (see Figure~\ref{fig:jet_signature}, bottom left).
As a first result, we confirm that,
\textit{within the B1 structure,} the highest velocity SiO emission (see the channel map centred at --10.70 km s$^{-1}$) is associated with the B1a clump \citep{benedettini07}.
This position is indeed considered the youngest impact between the jet (never directly imaged outside the protostar; \citealt{podio16})
and the B1 cavity (see also \citealt{busquet14} and \citealt{benedettini12} on
H$_2$O, OI and high-$J$ CO PACS Herschel data). 

Furthermore, the present dataset allows us to image, for the first time, very high velocity SiO emission outside B1, namely towards the northern B0 clump (see the channel map centred at --19.72 km s$^{-1}$ in Figure~\ref{fig:jet_signature}). 
The comparison between the spectra extracted in the three different intensity peaks clearly shows how the low velocity emission looses its importance in the B1a and B0 peaks.
The green spectrum in Figure~\ref{fig:jet_signature} shows a peak close to the systemic velocity of the source (+2.6 km s$^{-1}$, shown as a vertical line) and a very broad blue-shifted wing, with a hint of a second peak around --7.5 km s$^{-1}$. The orange spectrum shows a double peak structure with the blue-shifted peak brighter than the peak close to the systemic velocity.
The most extreme is represented by the blue spectrum, extracted at the peak position of the --19.72 km s$^{-1}$ channel map, which definitely differs with respect to the other two spectra, showing almost no emission at the systemic velocity and 
the expected profile for a shock driven by a high-speed jet (see e.g. \citealt{lefloch15}, and references therein). As B0 is the youngest shock in the region, it provides a unique opportunity to study which grain species are released by the shock to the gas phase.

\subsection{High and extremely high velocity SiO emission maps}\label{sec:LV-HV-EHV}
In order to better characterise the origin and nature of the SiO emitting gas, we roughly divided the integrated emission in three intervals, based on the spectra showed in Figure \ref{fig:jet_signature} and discussed in the previous section: low velocity (LV) from -3 to +5 km s$^{-1}$, high velocity (HV) from -13 to -3 km s$^{-1}$, and extreme high velocity (EHV) from -23 to -13 km s$^{-1}$. 

\begin{figure*}
\centering
 \includegraphics [width=1\textwidth]{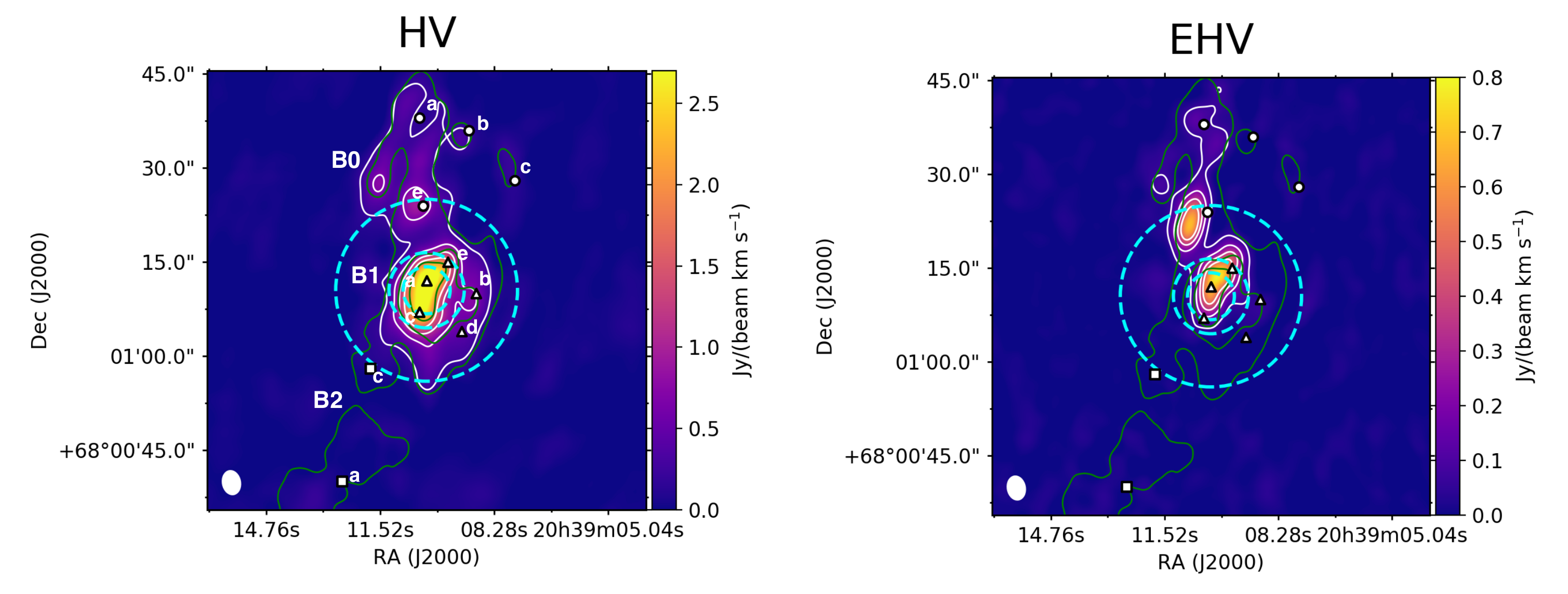}
 \caption{$^{28}$SiO emission integrated over two velocity ranges (see text): the extreme high velocity (EHV: --23 to --13 km s$^{-1}$) and the high velocity (HV: --13 to --3 km s$^{-1}$). The white contours start at 5$\sigma$, with steps of 15$\sigma$ (rms = 14 and 45 mJy beam$^{-1}$ for the EHV and HV, respectively). The subclumps composing the B0 and B1 shock regions along the outflow of L1157 are marked here with circles (B0), triangles (B1), and squares (B2). The dashed circles represent the  HPBW of the IRAM 30m spectra at 86 GHz (29\arcsec), 210 GHz (12\arcsec) and 350 GHz (7.5\arcsec). The green contours in the channel maps show the 10, 60 and 120 $\sigma$ levels of the integrated intensity map of the $^{28}$SiO, showed in Figure~\ref{fig:integrated-int}.
}
  \label{fig:lvg_emission size}
\end{figure*}

Figure \ref{fig:lvg_emission size} shows the emission maps of the HV and EHV gas, integrated in the corresponding velocity intervals, and Table \ref{tab:lines} reports the integrated intensities in those velocity intervals for all the SiO lines detected with single-dish telescopes. The single dish spectra are taken within the area marked by the circles in Figure  \ref{fig:lvg_emission size}.

The HV emission peaks at the B1a position, and is intense over a roundish clump more than 15$''$ in size. However, the HV emission is not limited to the B1a clump but originates also from the more extended B1 whole region as well as the B0 shock and walls.
On the other hand, the EHV emission is more concentrated and presents two distinct peaks, at the B1a and B0e positions, of similar intensity and spatial extent, about 15$''$ x 6$''$. A faint, but definitively detected, EHV emission is also present along the north wall of B0, perhaps identifying the gas entrained by the jet impacting B0 and responsible for the SiO emission shocked-gas clump in that position (and discussed in Section \ref{jet}).
In the next Section, we will analyse the gas properties (temperature and density) of the HV and EHV gas. 

\section{LVG analysis}\label{sec:lvg}
\begin{figure*}
\centering
 \includegraphics [width=1.0\textwidth]{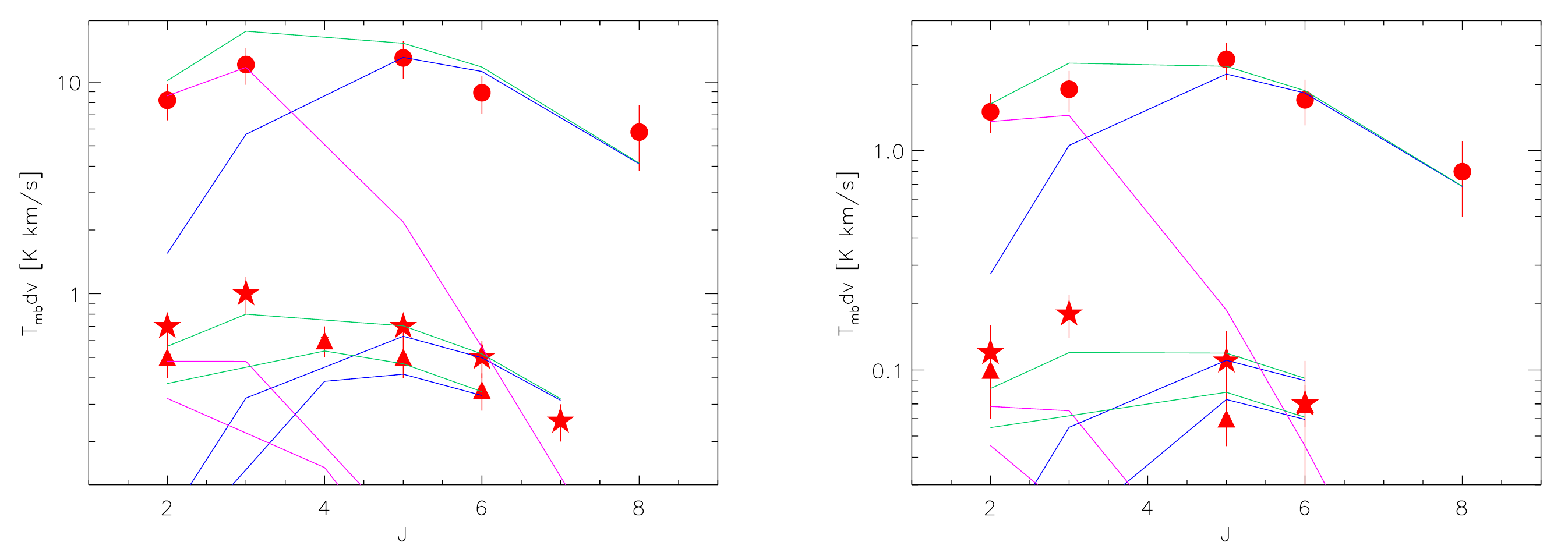}
 \caption{Result of the simultaneous fit of the $^{28}$SiO (circles), $^{29}$SiO (stars) and $^{30}$SiO (triangles) fluxes of the HV  (left panel) and EHV (right panel) components (Tab. \ref{tab:lines}), as a function of the upper level J of the transition. The HV best fit is obtained with two components: (i) a cold component (magenta curves) with $T$=40 K,  $n_{H2}=2\times10^{5}$ cm$^{-3}$, $N_{SiO}=4\times10^{13}$ cm$^{-2}$ and extended 25$''$, and (ii) a warm component associated with B1a (blue curves) with $T$=180 K, $n_{H2}=6\times10^{5}$ cm$^{-3}$, $N_{SiO}=4\times10^{13}$ cm$^{-2}$ and sizes=10$''$. The sum of the two components (green curves) fits extremely well the observations. 
 Similarly, the EHV best fit is obtained with two components: (i) a cold component (magenta curves) with $T$=30 K,  $n_{H2}=2\times10^{5}$ cm$^{-3}$, $N_{SiO}=8\times10^{12}$ cm$^{-2}$ and extended 20$''$, and (ii) a warm component associated with B1a (blue curves) with $T$=180 K, $n_{H2}=6\times10^{5}$ cm$^{-3}$, $N_{SiO}=8\times10^{12}$ cm$^{-2}$ and sizes=9$''$. The sum of the two components (green curves) fits the observations extremely well. 
}
  \label{fig:lvg}
\end{figure*}

In this section we analyse the single-dish spectra of the SiO and its isotopologues $^{29}$SiO and $^{30}$SiO from the $2-1$ up to the $8-7$ transition ($E_{\rm up}$ from $6$ to $163$ K), shown in Fig. \ref{fig:spectra}. The 12-11 spectrum has been excluded from the analysis because it is too noisy: note however that the fluxes reported in Table 1 are consistent with the solutions provided below.
The line properties are summarised in Table 1, which also reports the integrated intensity in three velocity ranges: extreme high velocity (EHV: --23 to --13 km s$^{-1}$), high velocity (HV: --13 to --3 km s$^{-1}$), and low velocity (LV: --3 to +5 km s$^{-1}$). 

In order to derive the average physical properties of the gas emitting SiO, we carried out a non-LTE analysis using the Large Velocity Gradient (LVG) code described in Ceccarelli et al. (2003). We used the collisional coefficients of SiO with H$_2$ computed between 10 and 300 K for the first 20 levels by \cite{dayou} and provided by the BASECOL database (Dubernet et al. 2013).

We run a large grid of models ($\geq$7000) covering a SiO column density $N_{SiO}$ from $1.5\times10^{13}$ to $8\times10^{15}$ cm$^{-2}$, a H$_2$ density $n_{H2}$ from $1\times10^{3}$ to $2\times10^{7}$ cm$^{-3}$, and a temperature $T$ from 30 to 300 K. We then find the solution by simultaneously fitting the $^{28}$SiO, $^{29}$SiO and $^{30}$SiO lines, leaving as free parameters $N_{SiO}$, $n_{H2}$ and $T$, while the emitting sizes and line widths are those measured. Note that we assumed the $^{28}$Si/$^{29}$Si=19.7 and $^{28}$Si/$^{30}$Si=29.8, as obtained by \cite{solar} and confirmed by our SOLIS observations for L1157-B1 (see  Section \ref{sec-fractionation}), and we added the 20\% of calibration uncertainty in the observed intensity error bars. We sampled the parameter space unevenly (with steps of 10 K until 100 K and 20 K for larger temperatures, steps of a factor of 2 for densities lower than $6\times10^6$ cm$^{-3}$ and 1.3 at larger ones, steps of a factor of 1.5 in column densities until $1\times 10^{14}$ cm$^{-2}$ and 1.1-1.2 at larger ones to be sure to cover it better close to the minimum $\chi^2$. 
The beam dilution is taken into account, assuming that the emitting gas has a circular gaussian shape and using the standard equation for beam dilution.

We carried out the non-LTE analysis for the HV and EHV components. The results are described in the two following subsections and summarised in Table \ref{tab:lvg-results}.

\begin{table}
 \caption{Result of the non-LTE LVG analysis. Both the HV and EHV emission is composed by two physical components: the first is associated with the B1a clump while the second is more extended (see text).}
     \scalebox{0.8}{
  \centering
  \begin{tabular}{lccccc}
    \hline
    \hline
    Component & $N_{SiO}$ & size & $T$ & $n_{H2}$ & X$_{SiO}$\tablefootmark{a} \\
              & $\times10^{12}$ cm$^{-2}$ & $''$ & K   & $\times10^{5}$ cm$^{-3}$ &$\times$10$^{-7}$\\
    \hline
    HV-B1a    & 40(8) & 10(2) & 180(36) & 6(1) & 5(1)\\
    HV-cold   & 40(8) & 25(5) &  40(8) & 2.0(4) &0.5(1)\\
    EHV-B1a   &  8(2) &  9(2) & 180(36) & 6(1) & 5(1)\\
    EHV-cold  &  8(2) & 20(4) &  30(6) & 2.0(4) & 0.5(1)\\
    \hline
 \end{tabular}}
  \tablefoot{
\tablefoottext{a}{SiO abundance with respect to molecular hydrogen. The H$_2$ column density used to calculate the SiO abundance is derived from the CO column density in the "g1" and "g2" components of the gas in L1157-B1 described in \cite{lefloch12}, see Section 5.3.}\\
{}{Numbers in parentheses are one standard deviation in units of the least significant digit.}\\
{}{The errors on the sizes are derived from the observations.}}
  \label{tab:lvg-results}
\end{table}

\subsection{High velocity component}

As shown in Figure \ref{fig:lvg_emission size} and discussed in Section \ref{sec:LV-HV-EHV}, the HV SiO emission extends over a size of 15$''$ centred on B1a with a halo encompassing the whole B1 region and the B0 wall. We, therefore, started our non-LTE analysis considering that the emission originates in the roundish 15$''$ B1a clump but could not find a good enough fit for all the J=2-1 to 8-7 $^{28}$SiO, $^{29}$SiO and $^{30}$SiO lines. When observing shocked regions like L1157-B1, the presence of several gas components in the line of sight has to be taken into consideration \cite{lefloch12}.
We then proceeded with a two physical components model, as suggested in the HV emission map by the presence of both the B1 and part of the B0 regions, see Figure \ref{fig:lvg_emission size}. To constrain the parameters of the gas around B1a, we used the three lines with a beam size approximately covering the B1a clump, namely those with J from 5-4 to 8-7.

The comparison between the observed and the theoretical integrated intensities provides a very good solution, with well constrained parameters: $N_{SiO}=$3--8$\times10^{13}$ cm$^{-2}$ for emitting size 10$"$ (i.e. approximately the observed FWHM of the HV emitting gas around B1a, shown in Fig. \ref{fig:lvg_emission size}), $n_{H2}$ between 10 and 4 $\times10^{5}$ cm$^{-3}$ and $T$ between 90 and 250 K, with the best fit $n_{H2}$--$T$ lying along a degenerate but very well constrained curve, where higher densities corresponds to lower temperatures.
We then looked for a fit with two physical components, the first one from the previous fit of the J=5-4 to 8-7 lines and the second one found by best fitting the residuals of all lines.
We obtain a final very good fit (reduced $\chi^2 =1.14$ with 11 degrees of freedom) with: (i) a warm component associated with the B1a clump with $T\sim180$ K, $n_{H2}\sim6\times10^{5}$ cm$^{-3}$, and $N_{SiO}\sim4\times10^{13}$ cm$^{-2}$; (ii) a colder component with $T\sim40$ K, $n_{H2}\sim2\times10^{5}$ cm$^{-3}$, $N_{SiO}\sim4\times10^{13}$ cm$^{-2}$ and whose extent is about $40''$x$15''$. The theoretical and the observed fluxes are shown in Fig. \ref{fig:lvg}. 
A similar good fit is also obtained considering for the warm B1a component $T$=100 K and $n_{H2}=1\times10^{6}$ cm$^{-3}$ ($\chi^2$=1.45) or $T$=250 K and $n_{H2}=4\times10^{6}$ cm$^{-3}$ ($\chi^2$=1.16). 

\subsection{Extremely high velocity component}
The SiO J=2-1 map of Figure \ref{fig:lvg_emission size} shows that the EHV emission is split in two clumps about 15$''$ x 6$''$ in size, centred on B1a and B0. We started with a single component fit with the emitting size equal to 18$''$, approximately the sum of the two clumps in the EHV map. We obtained a relatively good solution ($\chi^2$=1.8 for 9 degree of freedom) with $N_{SiO} =8\times10^{12}$ cm$^{-2}$, $n_{H2}$ between 1 and 7 $\times10^{5}$ cm$^{-3}$ and $T$ between 80 and 300 K, with the best fit $n_{H2}$--$T$ lying along a degenerate but very well constrained curve, where higher densities corresponds to lower temperatures.

However, there is no reason that the gas temperature and density are the same in the two clumps, so we also considered a fit with the lines encompassing the B1a clump only to have a more constrained view of the parameters in B1a. The best-fit of the J=5-4 to 8-7 $^{28}$SiO, $^{29}$SiO and $^{30}$SiO lines is obtained with $N_{SiO}$=8$\times10^{12}$ cm$^{-2}$, $n_{H2}=6\times10^{5}$ cm$^{-3}$ and $T=180$ K; within 1 $\sigma$, we obtain $N_{SiO}$ =4-40$\times10^{12}$ cm$^{-2}$, $n_{H2}$ between 3 and 10 $\times10^{5}$ cm$^{-3}$ and $T$ between 120 and 300 K, always along a degenerate curve as above.
We then proceeded, as for the HV case, to fit the residuals to constrain the parameters of a second physical component probed by the EHV emission. However, since the three lines (J=5 to 8) do not cover the B0 region, we only use the J=2-1 and 3-2 to constraint the EHV component in B0. With this caveat, we obtain a best fit of the second component with $N_{SiO}\sim 8\times10^{12}$ cm$^{-2}$, $n_{H2} \sim 2\times10^{5}$ cm$^{-3}$, $T \sim 30$ K and whose extent is about 30$''$x15$''$, indicating that the J=2-1 and 3-2 line emission is still dominated by the large scale encompassed by the single-dish measurements rather than the B0 clump. The obtained solution is shown in Figure \ref{fig:lvg}.

\section{Discussion}
\subsection{Isotopic fractionation and Galactic chemical evolution}
Isotopic fractionation is a key observable to understand the chemical evolution of the Universe. When studying the evolution of our Solar System, by observing the isotopic fractionation in some key elements we are able to distinguish among solar and pre-solar material, and hence derive information about the physical conditions of the pre-solar nebula, just before the formation of the Solar System (e.g. \citealt{hoppe18}). Moreover, isotopic fractionation can be used to get insights on star formation and evolution \citep{timmes96}.
\cite{monson17} show that secondary ($^{29}$Si and $^{30}$Si) to primary ($^{28}$Si) isotope ratios of silicon do not show any detectable variation along the Galactic radius. This is an interesting result because of the large ($\geq$900\%) variations observed in the secondary/primary oxygen isotope ratios, and Galactic chemical evolution (GCE) predicts that silicon and oxygen isotopic ratios should evolve in parallel with a nearly constant ratio in the secondary isotopes ($^{29}$Si/$^{30}$Si), and an increasing secondary to primary isotopic ratios ($^{29}$Si/$^{28}$Si and $^{30}$Si/$^{28}$Si) with decreasing Galactocentric radius \citep{wilson99}.
Furthermore, \cite{monson17} report a higher $^{30}$Si/$^{29}$Si in the ISM than both Solar and pre-solar SiC grains, hinting at the possibility of a difference in the chemical evolution of the two secondary isotopes of Si within the Galaxy. This result could be explained if a mass-dependent isotopic fractionation would be in place, where the heavier isotopes evolve differently with respect to the main one. We do not see this behaviour within L1157-B1, whose $^{29}$SiO/$^{30}$SiO corresponds within error bars to the Solar value across the whole source (see lower panel in Figure~\ref{fig:isotopic_ratio}). If a mass-dependent isotopic fractionation happens within our Galaxy, as suggested in \cite{monson17}, it does not occur in the timescales probed with L1157-B1 ($\sim$1000 years), i.e. the time scale that the material sputtered from the dust grains has to react in the gas phase and chemically evolve.

\subsection{The jet of L1157-mm}
Jets are pivotal in the early stages of star-formation, as they allow the accretion of mass from the disk onto the star by extracting angular momentum. While towards L1157 we see the effect of the jet impacting the gas \citep{tafalla95, gueth96, benedettini12}, the only direct observational evidence of the jet associated with L1157-mm so far are the spectra of the SiO bullets \citep{tafalla15} and the maps of the high velocity CO and SiO bullets \citep{podio16}, both detected close to the protostar. 
We present here for the first time a very high velocity bullet in SiO, showing the high-velocity jet at the position of the B0 shocked region working surface (at a distance of 68$''$, i.e. 0.12 pc, from the protostar). This SiO bullet is barely recognisable in the channel maps presented in Figure 2 of \cite{gueth98} (lower panel to the right, at -18.3 km s$^{-1}$). Moreover, the velocity found in the high-velocity SiO bullet B0 is compatible with the velocity of the high-velocity bullets detected close to L1157-mm in \cite{podio16} if we take into account the precession pattern. 
A high-velocity feature was already observed in CS (see red curve in Figure 9 of \citealt{benedettini13}) towards the same position.
The spectrum extracted from the peak of the high velocity channel (blue solid line) in Figure~\ref{fig:jet_signature} shows the signature of the jet impacting the gas and creating the shocked region B0. 
A similar spectral feature has been already observed towards L1448-B1 in \cite{nisini07}, see their spectrum at the bottom left in Figure 2, and towards IRAS 04166+2706 in \cite{tafalla10}, see top right spectrum in their Figure 3.

\subsection{The B1a clump and its surrounding gas}
The high spatial resolution of the new SOLIS SiO observations coupled with single-dish observations allow to provide stringent constraints on the gas shocked at the B1a position and its surrounding gas. The analysis and the derived gas parameters are discussed in Section \ref{sec:lvg} and summarised in Table \ref{tab:lvg-results}.

The SiO emitting gas in B1a, both the HV and EHV components, is heated at about 180 K and its density is about $6\times10^5$ cm$^{-3}$: there is about five times more SiO column density associated with the HV than the EHV gas in B1a. 
Similarly, the gas surrounding B1a is much colder, at around 30--40 K, and more tenuous by about a factor 3. The SiO column density of this cooler gas is the same as that in B1a, which is a bit surprising as a coincidence.
In the following, we will try to better understand the meaning of these findings and the implications.

A few years ago, \cite{lefloch12} carried out an analysis of the CO emission towards L1157-B1 using single-dish observations that covered the J=1-0 up to J=16-15 lines. They found that the spectra could be decomposed in three velocity components, defined by the slope of the intensity versus the velocity of each line spectrum. Although the velocity decomposition is different from that used in the present work, which is purely based on three different velocity intervals, it is interesting to note that Lefloch et al. also find a component, designed as "g1", which they associated with the shocked gas, whose sizes ($\sim 10''$), density (a few $10^6$ cm$^{-3}$) and temperature ($\sim 200$ K) is not very different from our component associated with B1a. Therefore, our new SOLIS observations suggest that the Lefloch et al. "g1" is very likely dominated by the emission from the shocked gas at B1a, supporting the Lefloch et al. claim. Incidentally, this indicates that an accurate non-LTE analysis of single-dish data can provide fairly correct information even at scales much smaller than the used telescope beams.

In addition to the "g1" component, Lefloch et al. also found two other components, called "g2" and "g3", associated respectively with the walls of the B1 and B2 cavities. The "g2" component has parameters not very different from those of the "cold" component of our HV and EHV fit: sizes about 20$''$ against our 20--25$''$, temperature about 60 K against our 30--40 K, and density larger than a few $10^5$ cm$^{-3}$ against our $\sim2\times10^5$ cm$^{-3}$. Therefore, the "g2" component is probably, at least in part, associated with the gas surrounding B1a and encompassing the whole B1 clump. 

The association of our B1a and cold components with the "g1" and "g2" components of the Lefloch et al. CO analysis allow us to estimate the abundance of SiO in them. Specifically, Lefloch et al. found that the CO column density is $\sim 1\times10^{16}$ and $\sim 1\times10^{17}$ cm$^{-2}$ in "g1" and "g2", respectively: assuming a CO abundance of $10^{-4}$ with respect to H$_2$, this corresponds to a SiO abundance of  about 5 and 0.5 $\times10^{-7}$ in B1a and its surrounding, respectively. Although substantially increased with respect to the SiO in the quiescent gas of molecular clouds ($\sim$10$^{-12}$, \citealt{martin-pintado92}), the silicon in the gas phase, released because of the sputtering and shattering of the grain mantles and refractory cores, is still just a small fraction of its elemental abundance ($\sim$3.5$\times10^{-5}$, \citealt{wilson94}). The remaining Si is likely still in the dust grains, as they are not expected to be completely destroyed by shocks. Assuming that atomic Si is efficiently converted into SiO as soon as it reaches the gas-phase \citep{neufeld89, herbst89}, our results reach a good agreement with the models of \cite{caselli97} for a shock with with velocity between 30 and 35 km s$^{-1}$. 

We finally discuss the nature/origin of the "cold" gas surrounding B1a.
Based on its properties (temperature, density and SiO abundance), this gas can be either material entrained by the B1a shock or itself cooled gas from previous (namely older) shocks or both. The presence of multiple shocks in B1, that occurred within 1500 yr, was reported by Podio et al. (2016) and successfully used by Codella et al. (2017, 2019) to identify chemical formation routes of molecules more complex than SiO, so there is a high probability that the "cold" component is cooled shocked gas. However, the HV map of the SiO J=2-1 line (Figure \ref{fig:lvg_emission size}) suggests that a more extended, probably entrained gas also contributes to the emission.

\section{Conclusions}
We present here the first isotopic ratio maps of SiO towards a shocked region.
From these interferometric maps we can infer that the (2-1) transition of the main isotopologue is optically thick within the whole region imaged, also at high velocities (up to -20 km s$^{-1}$). The isotopic ratio map of the two rarer isotopologues, $^{29}$SiO and $^{30}$SiO, shows a uniform value consistent with the Solar value of [$^{29}$Si/$^{30}$Si]= 1.5.

Coupling the high spatial resolution of the new SOLIS data with the large single-dish dataset, we were able to put stringent constraints on the physical properties of the gas in L1157-B1. Our observations can be reproduced by a more compact, denser and warmer gas component ($n_{H2}$ = 6$\times$10$^{5}$ cm$^{-3}$, T = 180 K and X$_{Si}$ = 5$\times$10$^{-7}$), and a more extended colder and less dense gas component ($n_{H2}$ = 2$\times$10$^{5}$ cm$^{-3}$, T = 40 K and X$_{Si}$ = 0.5$\times$10$^{-7}$).

Furthermore, thanks to the high spectral resolution and sensitivity of the SOLIS dataset, we have identified a high-velocity bullet of SiO tracing the impact of the jet arising from the protostar L1157-mm on the B0 clump, opening a new laboratory for future studies to further unveil the complex physical and chemical structure of this source.

\begin{acknowledgements}
We thank the anonymous referee for valuable comments and suggestions. This work is based on observations carried out under project number L15AA with the IRAM NOEMA Interferometer [30m telescope]. IRAM is supported by INSU/CNRS (France), MPG (Germany) and IGN (Spain). We are very grateful to all the IRAM staff, whose dedication allowed us to carry out the SOLIS project. This work was supported by (i) the PRIN-INAF 2016 "The Cradle of Life - GENESIS-SKA (General Conditions in Early Planetary Systems for the rise of life with SKA)", (ii)  the program PRIN-MIUR 2015 STARS in the CAOS - Simulation Tools for Astrochemical Reactivity and Spectroscopy in the Cyberinfrastructure for Astrochemical Organic Species (2015F59J3R, MIUR Ministero dell'Istruzione, dell'Universit\'a della Ricerca e della Scuola Normale Superiore), (iii) the European Research Council (ERC) under the European Union's Horizon 2020 research and innovation programme, for the Project "The Dawn of Organic Chemistry" (DOC), grant agreement No 741002, and (iv) the European Marie Sk\l{}odowska-Curie actions under the European Union's Horizon 2020 research and innovation programme, for the Project "Astro-Chemistry Origins" (ACO), Grant No 811312.
\end{acknowledgements}

{}

\begin{appendix}
\section{Channel maps}
Figures A.1, A.2 and A.3 show the channel maps of the observed SiO isotopologues (see Table 1) with low velocity resolution (7.3 km s$^{-1}$). Figures A.4 and A.5 show the channel maps of the main isotopologue observed with high velocity resolution (0.5 km s$^{-1}$).

\begin{figure*}
\centering
 \includegraphics [width=0.8\textwidth]{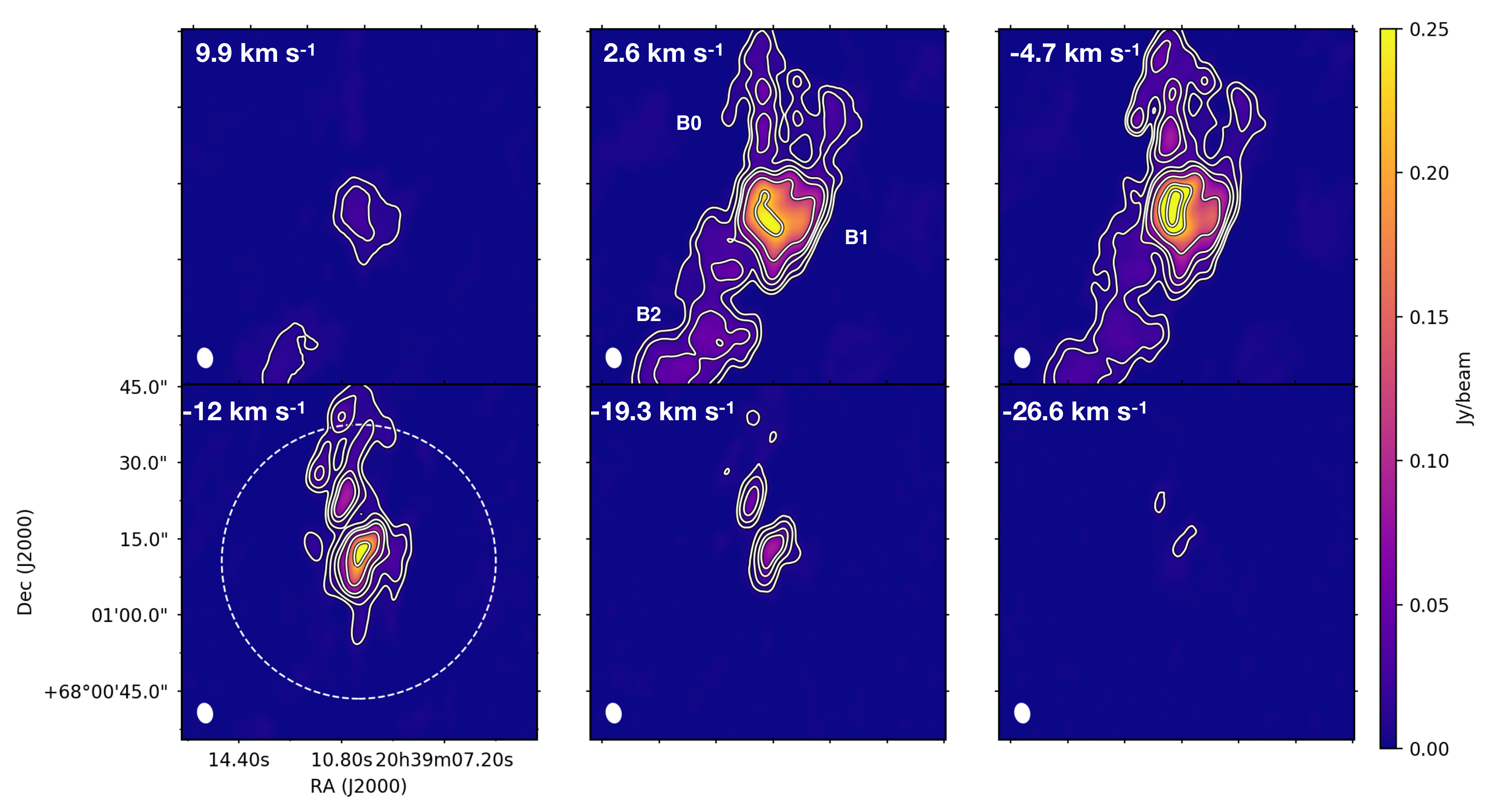}
 \caption{ $^{28}$SiO channel maps observed with a spectral resolution of 7.3 km s$^{-1}$. The clean beam of 3.8$\arcsec$ $\times$ 2.8$\arcsec$ is shown at the bottom left of the maps, and the primary beam of 56$\arcsec$ is reported as a dashed white circle, in the bottom left panel. The contours are 10, 20, 40 and 60, 120, and 250 and 350 $\sigma$ (rms = 0.9 mJy/beam).
}
\end{figure*}

\begin{figure*}
\centering
 \includegraphics [width=0.8\textwidth]{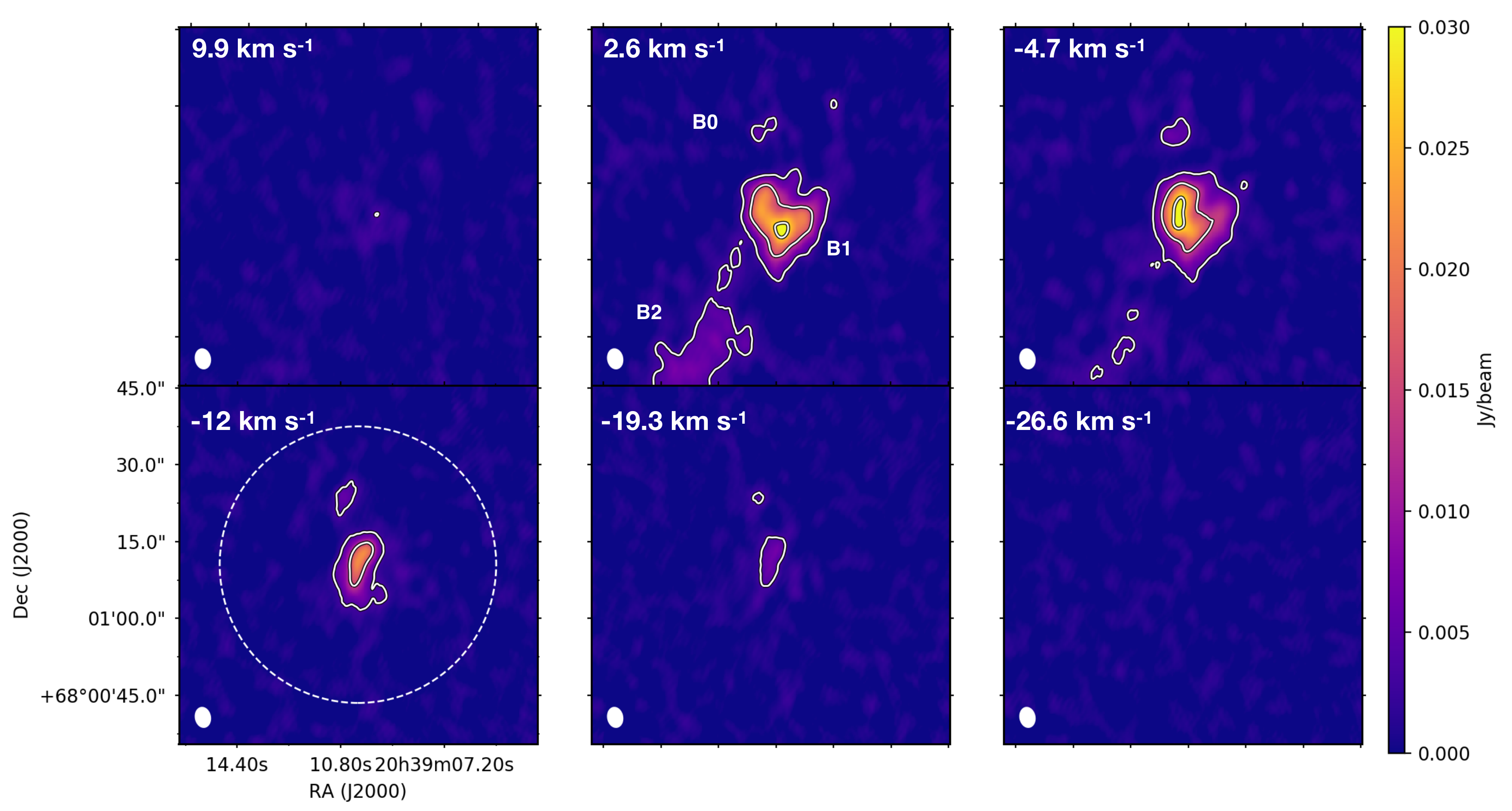}
 \caption{ $^{29}$SiO channel maps observed with a spectral resolution of 7.3 km s$^{-1}$. The clean beam of 3.8$\arcsec$ $\times$ 2.8$\arcsec$ is shown at the bottom left of the maps, and the primary beam of 56$\arcsec$ is reported as a dashed white circle, in the bottom left panel. The contours are 5, 20 and 40 $\sigma$ (rms = 0.7 mJy/beam).
}
\end{figure*}

\begin{figure*}
\centering
 \includegraphics [width=0.8\textwidth]{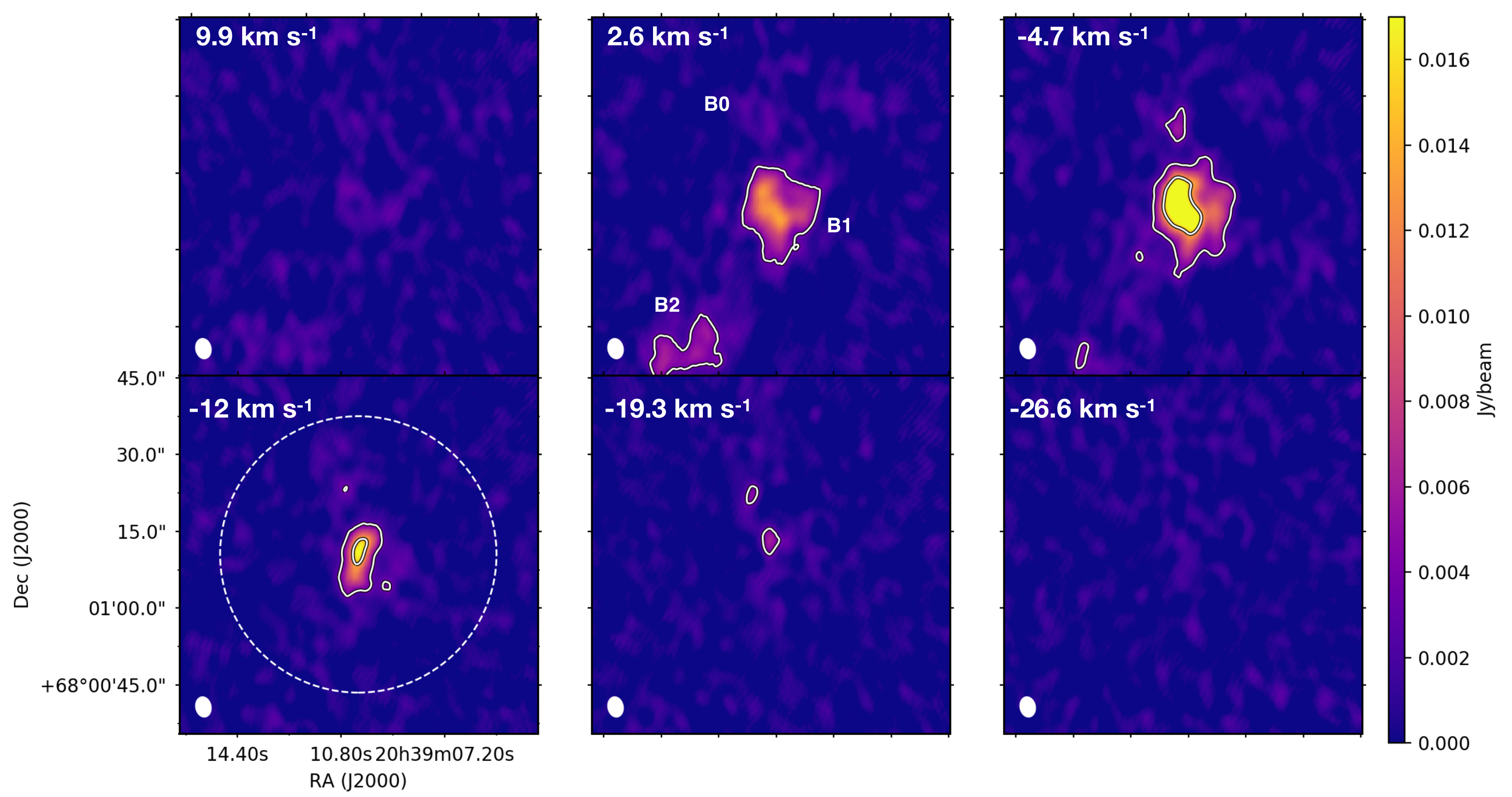}
 \caption{$^{30}$SiO channel maps observed with a spectral resolution of 7.3 km s$^{-1}$. The clean beam of 3.8$\arcsec$ $\times$ 2.8$\arcsec$ is shown at the bottom left of the maps, and the primary beam of 56$\arcsec$ is reported as a dashed white circle, in the bottom left panel. The contours are 5 and 20 $\sigma$ (rms = 0.7 mJy/beam).
}
\end{figure*}

\begin{figure*}
\centering
 \includegraphics [width=1\textwidth]{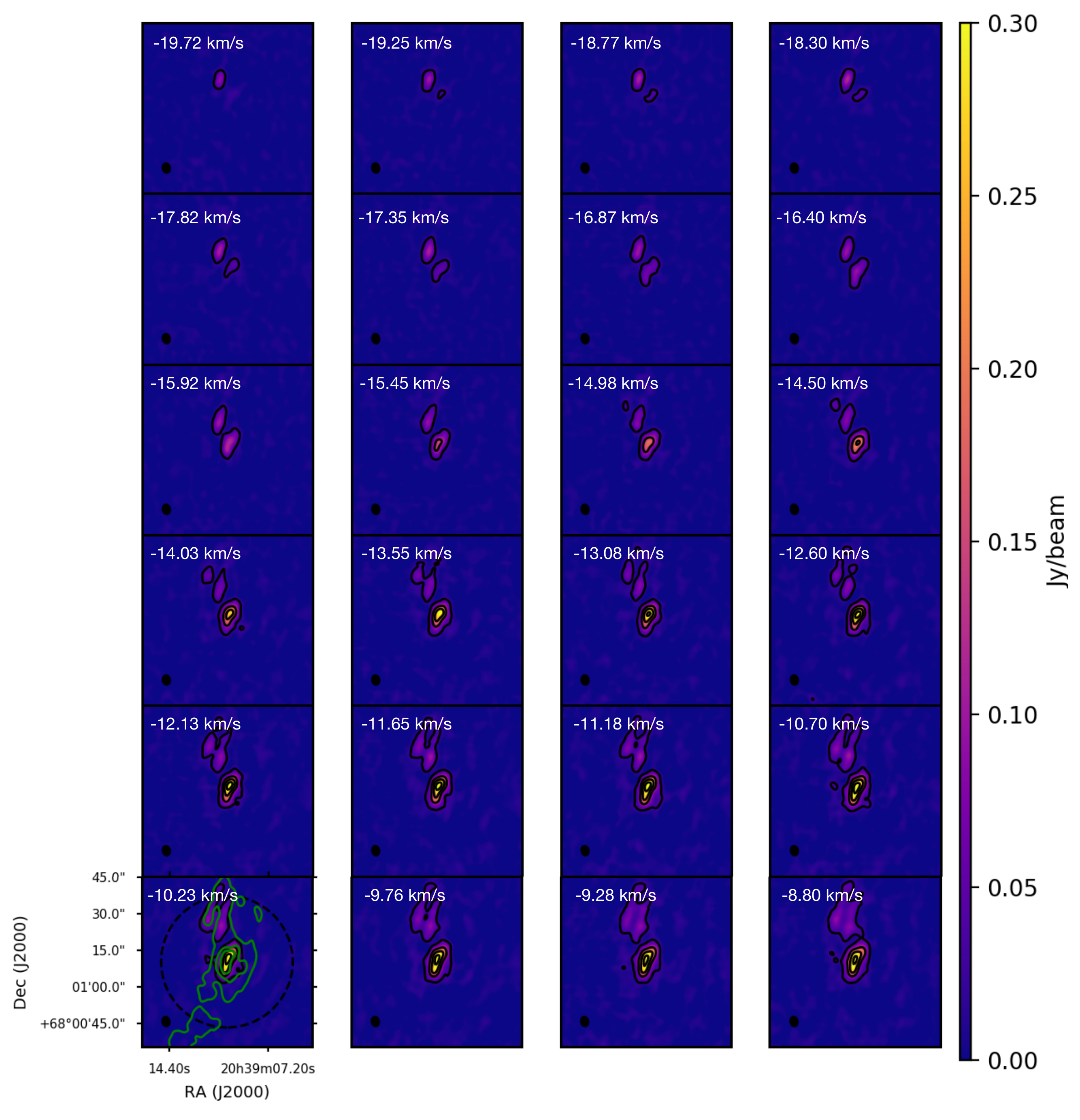}
 \caption{$^{28}$SiO channel maps observed with a spectral resolution of $\sim$ 0.5 km s$^{-1}$. The clean beam of 3.8$\arcsec$ $\times$ 2.8$\arcsec$ is shown at the bottom left of the maps, and the primary beam of 56$\arcsec$ is reported as a dashed black circle. The green contours in the channel maps show the 10, 60 and 120 $\sigma$ levels of the integrated intensity map of the $^{28}$SiO, showed in Figure~\ref{fig:integrated-int}. The central velocity of the channel is reported in the top left corner of each panel.
}
\end{figure*}

\begin{figure*}
\centering
 \includegraphics [width=1\textwidth]{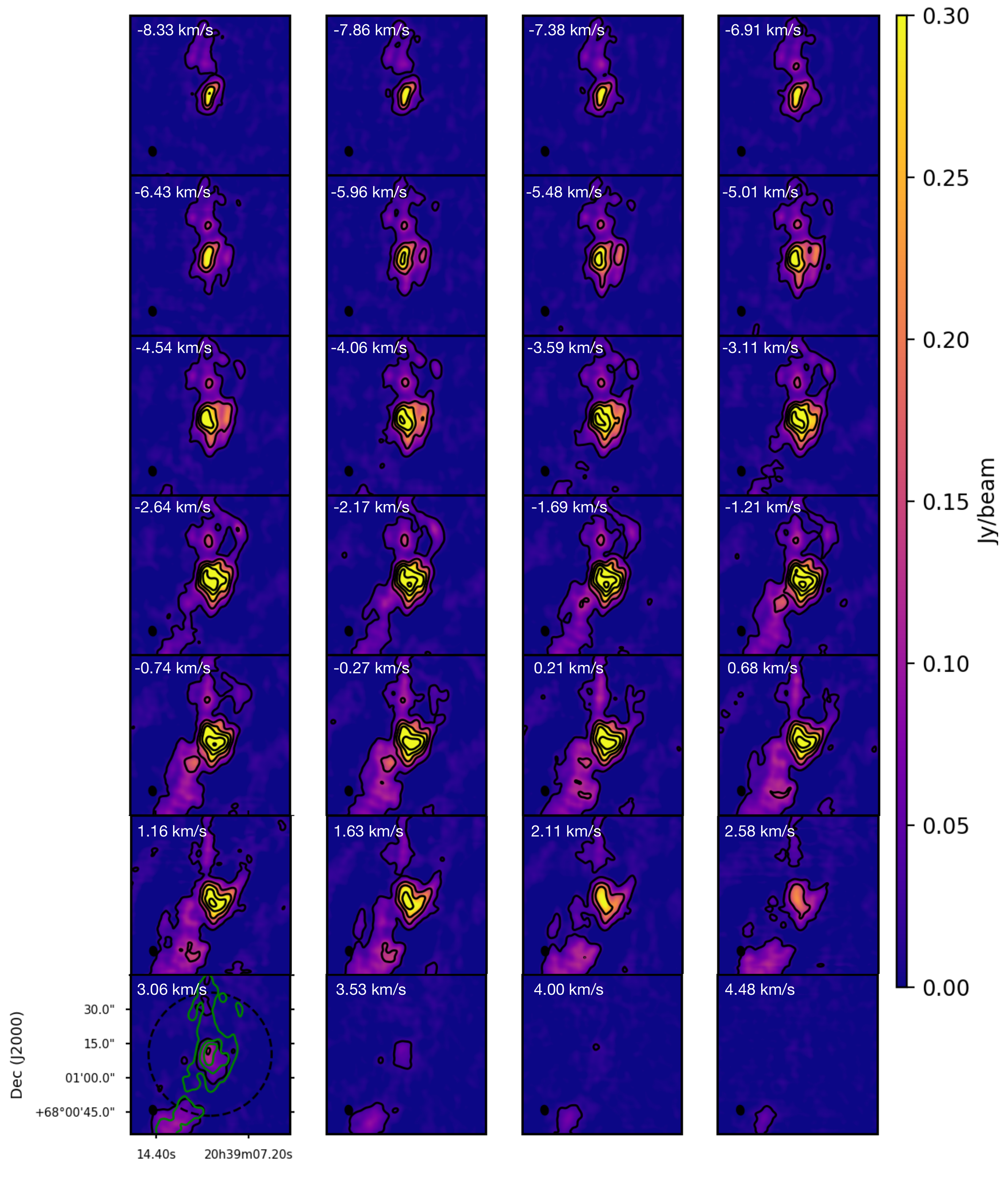}
 \caption{ continue as Figure A.4
}
\end{figure*}

\section{Missing flux}
The spectra of the main isotopologue $^{28}$SiO observed with NOEMA has been compared with the spectrum observed with the IRAM 30-m antenna, to evaluate the missing flux
of the present NOEMA dataset. The comparison is shown in Figure B.1.
About 80\% of the flux is recovered.

\begin{figure*}
\centering
 \includegraphics [width=0.5\textwidth]{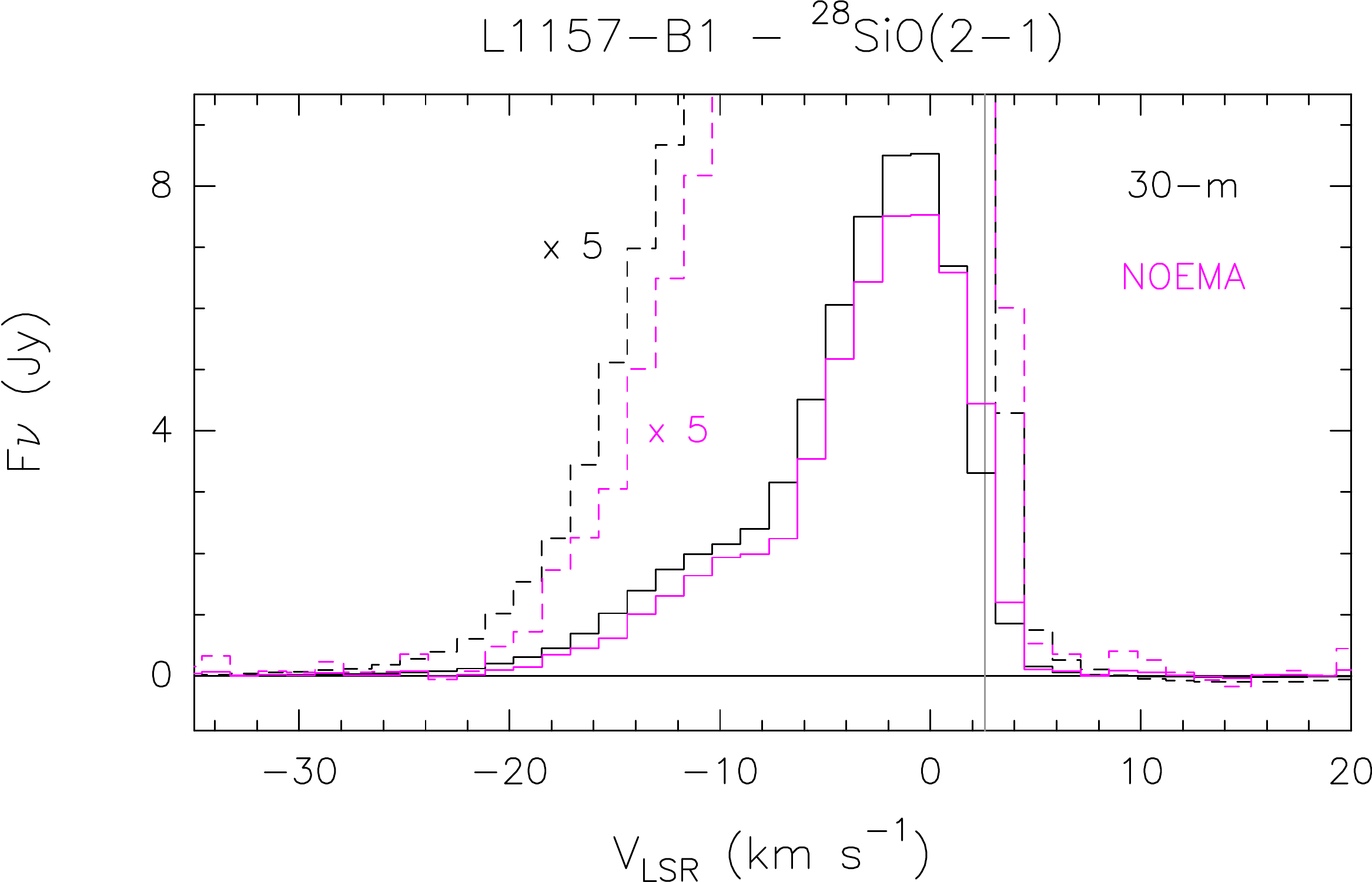}
 \caption{$^{28}$SiO 2-1 intensity profile extracted from the data cube averaged within a 30$\arcsec$ beam (magenta line) is over-plotted on the IRAM-30m spectrum (HPBW $\sim$ 30$\arcsec$) (black histogram).}
\end{figure*}

\end{appendix}

\end{document}